# Stabilization of Martensite and Austenite Phases and Realization of Two-way Martensitic Transition in Co-Ni-Ga Ferromagnetic Shape Memory Alloy Nanoparticles

*Debraj Mahata[a], and Ananthakrishnan Srinivasan[a]\**

[a] Department of Physics, Indian Institute of Technology Guwahati, Guwahati – 7381039, India.

ABSTRACT

Three sets of Co-Ni-Ga alloy nanoparticles have been synthesized by a template-free chemical route. Structural, morphological, shape memory, and magnetic properties of room temperature martensite (M) phase, dual (M + secondary γ) phase and austenite (A) phase Co-Ni-Ga nanoparticles are reported. Temperature-dependent XRD analysis revealed that $Co_{36}Ni_{36}Ga_{28}$ nanoparticles exhibiting a single M phase at room temperature, completely transform to A phase at ~1000 K. Upon cooling to room temperature, the A phase transforms back to single M phase, confirming the two-way martensitic transition in Co-Ni-Ga nanoparticles. Structural analysis shows that the γ-phase does not influence the martensitic transition of bi-phasic (M + γ) $Co_{41}Ni_{34}Ga_{25}$ nanoparticles. These nanoparticles display saturation magnetization ranging from 2.9 emu/g to 15.3 emu/g at room temperature. The γ phase could be introduced in A phase $Co_{44}Ni_{26}Ga_{30}$ nanoparticles when heated up to 1073 K. Curie temperatures of A and M phases are higher than the martensitic transition temperatures in all the samples, qualifying them as ferromagnetic shape memory alloy nanoparticles. Observation of M ↔ A phase transition, Co-Ni-Ga nanoparticles with tunable magnetic properties make them excellent candidates for low and high temperature nanoactuators and other ferromagnetic shape memory applications.

1. INTRODUCTION

Shape memory nanoparticles are gaining attention due to the recent trend of miniaturizing functional devices based on the shape memory effect [1-3]. While bulk shape memory alloys (SMAs) have been extensively studied, their nanoparticle counterparts which have the potential to harness quantum confinement effects, remain largely unexplored. It has been reported that the martensitic transition can get suppressed in SMA nanoparticles [4,5], although a few attempts have been made to study the martensitic transition in smaller particles [1-3,6]. Though metallurgical methods, like recrystallization from amorphous solid solutions, are commonly used to produce nanometer sized SMA phases [6-8], others including ball milling [9], laser-induced pyrolysis [10], and vapor deposition [11] have also been explored. Recently, use of template has become a popular method for chemically synthesizing Heusler alloy nanoparticles with a controlled size distribution [12-13]. While this method has greatly advanced strain-free phase stabilization of Heusler alloy nanoparticles, the templates remain as impurities along with the desired alloy phase, influencing their structural as well as magnetic properties. Recently, Co and Fe-based Heusler alloy nanoparticles were synthesized successfully using template-free chemical methods [14-16].

Ni-Mn-based alloys [17-19] have become the leading and most studied ferromagnetic SMAs (FSMAs), following the groundbreaking discovery of a magnetic field-induced significant strain in Ni-Mn-Ga alloy [20-22]. However, low martensitic transition temperature and high brittleness limit the practical use of Ni-Mn-Ga alloy [23-24]. These limitations can be addressed by enhancing the ductility through the addition of a secondary γ phase in Co-Ni-Ga FSMA [25-27]. The γ phase significantly influences the mechanical properties along with the magnetic properties of Co-Ni-Ga FSMA [13,27]. Thus, the synthesis of impurity-free Co-Ni-Ga nanoparticles with an appropriate amount of γ phase remains an intriguing and compelling challenge for researchers. In 2015, Wang *et al.* reported the synthesis of γ-Co-Ni-Ga nanoparticles through a colloidal silica template-assisted chemical method [13]. Subsequently, template-assisted chemical methods were used to obtain biphasic dual phase (A + M and A + γ) Co-Ni-Ga nanoparticles [5, 12,13,28]. In 2024, our group reported a procedure to individually stabilize A, M, γ, A + M, M + γ, and A + M + γ phases in chemically synthesized Co-Ni-Ga nanoparticles [29]. However, no studies have been reported on the martensitic transition in Co-Ni-Ga nanoparticles amid speculations that the martensitic transition might get smeared out in fine particles of the FSMA [4]. Moreover, the effects of γ phase on the mechanical and magnetic properties, which are well documented in bulk Co-Ni-Ga alloy, have not been explored in Co-Ni-Ga nanoparticles. The limited research on martensitic transition in Co-Ni-Ga alloy nanoparticles inspired us to investigate these nanoparticles synthesized using a facile template free chemical method.

This report explores the structural, morphological, and magnetic properties of three sets of Co-Ni-Ga nanoparticles prepared through a template-free chemical route with single A, single M, and dual (M + γ) phases. This study also examines the martensitic transition occurring at high and low temperatures and the impact of the γ phase on the martensitic transition in the nanoparticles.

2. EXPERIMENTAL

Co-Ni-Ga nanoparticles were synthesized using high (99.9%) purity $CoCl_2·6H_2O$, $Ni(NO_3)_3·6H_2O$, and $Ga(NO_3)_3·xH_2O$ metal precursors, procured from Sigma-Aldrich or Alfa Aesar. The metal precursors were added in right proportion to methanol and subjected to 30 minutes of sonication to ensure thorough mixing. The thoroughly mixed solution was heated at 80 °C under constant stirring. Upon complete drying, the samples underwent heat treatment at designated temperatures for 3 hours each in a tubular furnace under continuous purging of $N_2$ and $H_2$ gases. Following heat treatment, the samples were gently ground into a fine powder for characterization. The specific heat treatment conditions and the quantities of precursors utilized to prepare each sample are detailed in Table 1.

**Table 1.** Amount of precursors and heat treatment conditions used to synthesize various Co-Ni-Ga (CNG) nanoparticles.

| Sample id | $CoCl_2.6H_2O$ (g) | $Ni(NO_3)_3.6H_2O$ (g) | $Ga(NO_3)_3.xH_2O$ (g) | Temperature (°C) | Time (hours) |
|---|---|---|---|---|---|
| CNG-1 | 0.7000 | 0.8000 | 1.2536 | 950 | 3 |
| CNG-2 | 0.7217 | 0.8033 | 1.2675 | 950 | 3 |
| CNG-3 | 0.8284 | 0.5574 | 1.3577 | 750 | 3 |

Room temperature crystal structure of the Co-Ni-Ga nanoparticles was determined using a powder X-ray diffractometer (Rigaku Smartlab) operating in Bragg-Brentano geometry with an input power of 5 kW. X-ray diffraction (XRD) patterns were recorded using a Phillips X'Pert Pro X-ray diffractometer equipped with a high temperature stage operating with Cu-K$_\alpha$ radiation with a wavelength of 1.5406 Å. A field emission scanning electron microscope (FESEM, Zeiss Sigma 300) equipped with an energy dispersive spectrometer (EDS, Zeiss Sigma) was utilized to assess the elemental composition and examine the surface morphology of the synthesized nanoparticles. A vibrating sample magnetometer based on an electromagnet (VSM, Lakeshore, 7410) and another incorporated in a physical properties measurement system (VSM-PPMS, Quantum Design, Dynacool) were employed for measuring the magnetic properties.

3. RESULTS AND DISCUSSION

3.1. Crystal structure

Figure 1 shows the room temperature XRD patterns of all three chemically synthesized Co-Ni-Ga nanoparticles. The XRD patterns confirm the single M phase in CNG-1, M + γ dual phase in CNG-2 and single A phase in CNG-3 at room temperature. The M phase has a body centred tetragonal structure with space group 123, *P*/4*mmm* (ICSD file no. 169730) and Wyckoff positions Co (0 0 0) and Ni/Ga (½ ½ ½). The nanoparticles (CNG-1 and CNG-2) were sprinkled on the Pt plate of the high temperature sample holder for recording the XRD pattern. Due to the exposure of the Pt plate to the X-rays, extra peaks around 2$\theta$ of 40°, 45°, 68°, and 82° arise from FCC Pt with lattice constant $a$ = 3.918 Å. EDS data confirms the overall composition of single M phase in CNG-1 nanoparticles to be Co:Ni:Ga ≈ 36:36:28 (at.%). Starting with the composition of CNG-1, when its Co content is increased to 41 at.% at the expense of Ni (34 at.%) and Ga (25 at.%), some amount (~19%) of γ phase appeared along with the M phase as indicated by the room-temperature XRD pattern of CNG-2.

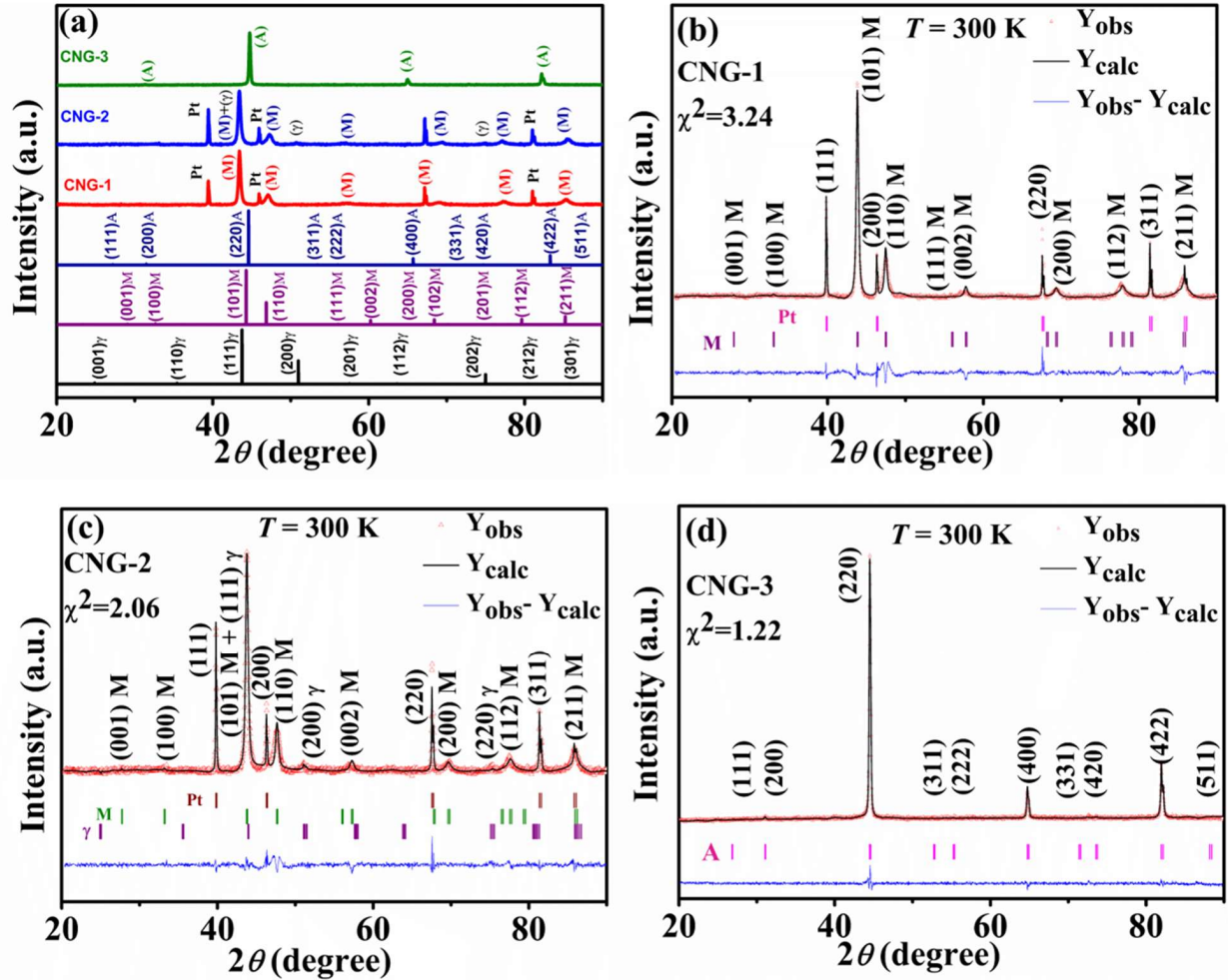

**Figure. 1** (a) Room temperature experimental and simulated XRD patterns of the three Co-Ni-Ga nanoparticles. Simulated patterns have been generated using unit cells of A, M, and γ phases. Rietveld refined XRD patterns of (b) CNG-1, (c) CNG-2, and (d) CNG-3 nanoparticles.

Earlier studies have generally identified the secondary γ phase as an FCC structure with a lattice parameter of ~3.60 Å [30]. However, Wang *et al.* [13] carefully analyzed the XRD pattern along with SAED pattern of γ-Co-Ni-Ga nanoparticles and showed their structure is more appropriately described by a cubic structure with slight tetragonal distortion (space group: 123, *P*4/*mmm*, ICSD file no: 157788), resembling the prototype $Pt_2FeCu$ compound. SAED pattern analysis of our synthesized M+γ-Co-Ni-Ga (CNG-2) nanoparticles (not shown here) confirms the tetragonal distorted cubic structure proposed for the γ-Co-Ni-Ga phase by Wang *et al.* Such dual M + γ phase has also been reported in Co-Ni-Ga shape memory alloy [31,32].

**Table 2.** Elemental composition and structural properties of CNG-1, CNG-2, and CNG-3.

| Sample id | Composition Co:Ni:Ga (at.%) | A phase | | M phase | | | γ phase | | | $\chi^2$ |
|---|---|---|---|---|---|---|---|---|---|---|
| | | $a$ (Å) | Phase % | $a$ (Å) | $c$ (Å) | Phase % | $a$ (Å) | $c$ (Å) | Phase% | |
| **CNG-1** | 36:36:28 | --- | --- | 2.709(3) | 3.192(3) | 100 | --- | --- | --- | 3.24 |
| **CNG-2** | 41:34:25 | --- | --- | 2.698(5) | 3.216(8) | 81 | 3.574(9) | 3.553(8) | 19 | 2.06 |
| **CNG-3** | 44:26:30 | 5.754(4) | 100 | --- | ---- | --- | --- | --- | --- | 1.22 |

Pure A phase can be achieved by increasing the Co content further to 44 at.% and Ga to 30 at.% and decreasing Ni to 25 at.%. The room temperature XRD pattern confirms the phase purity of the A phase CNG-3 nanoparticles. *B*2 type of partially disordered Heusler alloy structure (space group: 225, $Fm\bar{3}m$ (ICSD file no. 169731) has been assigned for CNG-3 with a single A phase at room temperature.

Table 2 presents the unit cell parameters, phase percentages, and goodness of Rietveld fit parameter ($\chi^2$) of CNG-1 (M phase), CNG-2 (dual M + γ phase), and CNG-3 (A phase) nanoparticles as obtained from the Rietveld refinement process. The average crystallite size ($D_v$) was determined using Sch*ë*rrer's equation [33]. $D_v$ of CNG-1, CNG-2, and CNG-3 nanoparticles are 22 ± 1 nm, 20 ± 1 nm, and 25 ± 1 nm, respectively. The reported $D_v$ values for Co-Ni-Ga nanoparticles, synthesized via a template-assisted chemical approach, range between 12 nm and 69 nm [5,12,13].

## 3.2. Morphology

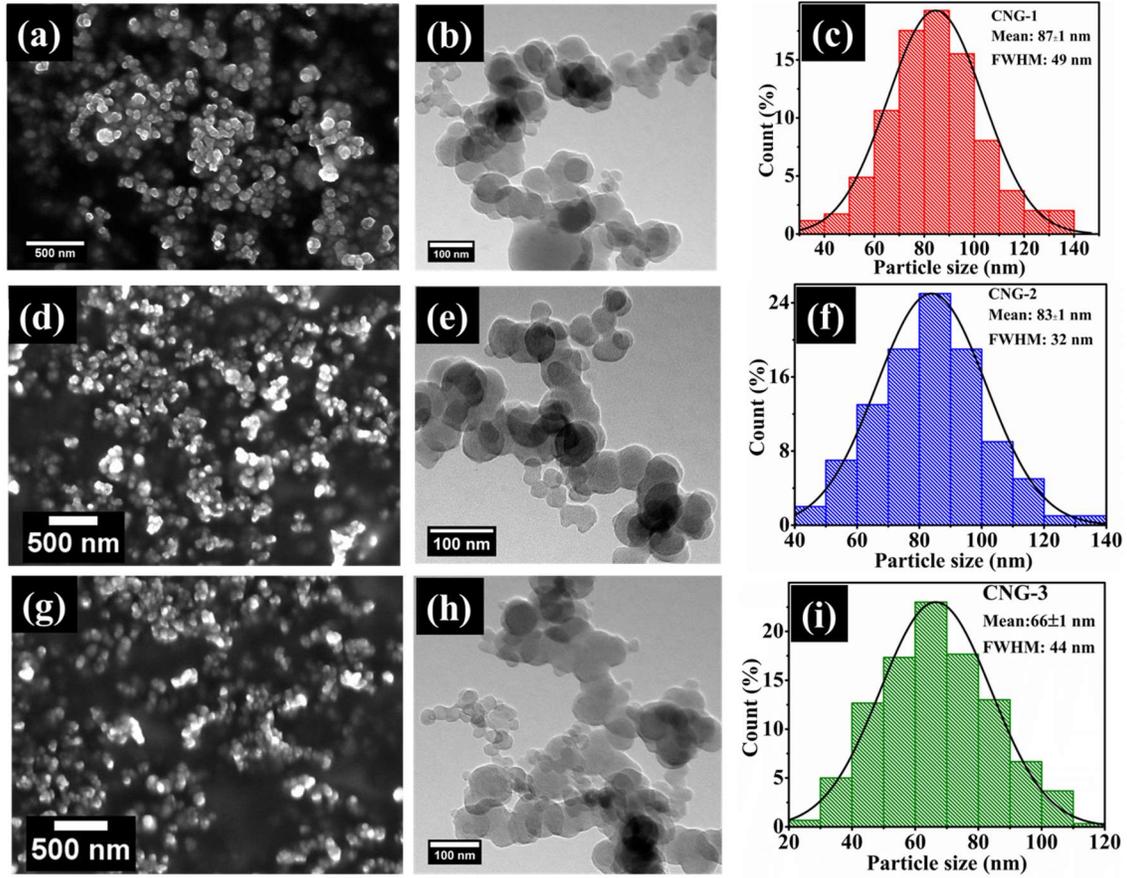

**Figure 2.** FESEM micrograph, FETEM image and particle size distribution of single M (CNG-1), dual M + γ (CNG-2), and single A (CNG-3) phase nanoparticles.

Figure 2 presents the FESEM and FETEM micrographs of CNG-1, CNG-2, and CNG-3 nanoparticles. The particle size distributions representing the corresponding FESEM micrograph are fitted to a Gaussian function as shown in Figure 2 (c, f, i). The average particle size of CNG-1, CNG-2, and CNG-3 nanoparticles is 87 ± 1 nm, 83 ± 1 nm, and 66 ± 1 nm, respectively. The corresponding full width at half maxima (FWHM) values of the normal distribution for CNG-1, CNG-2, and CNG-3 nanoparticles are 49, 32, and 44 nm, respectively. The figures clearly illustrate that the majority of the synthesized nanoparticles exhibit a spherical morphology. Similar structural characteristics have been documented in previous studies [34-35], which employed a template-free chemical method for the synthesis of other Heusler alloy nanoparticles. Prominent agglomeration and substantial FWHM observed in the nanoparticles can be ascribed to their intrinsic magnetic properties, coupled with the lack of any mechanism to physically separate the nanoparticles in the form a template. While notably high FWHM values can also be attained through template-assisted methods [5,13], the present FWHM values fall within acceptable values considering the advantage of the template-less approach.

3.3. Shape memory effect

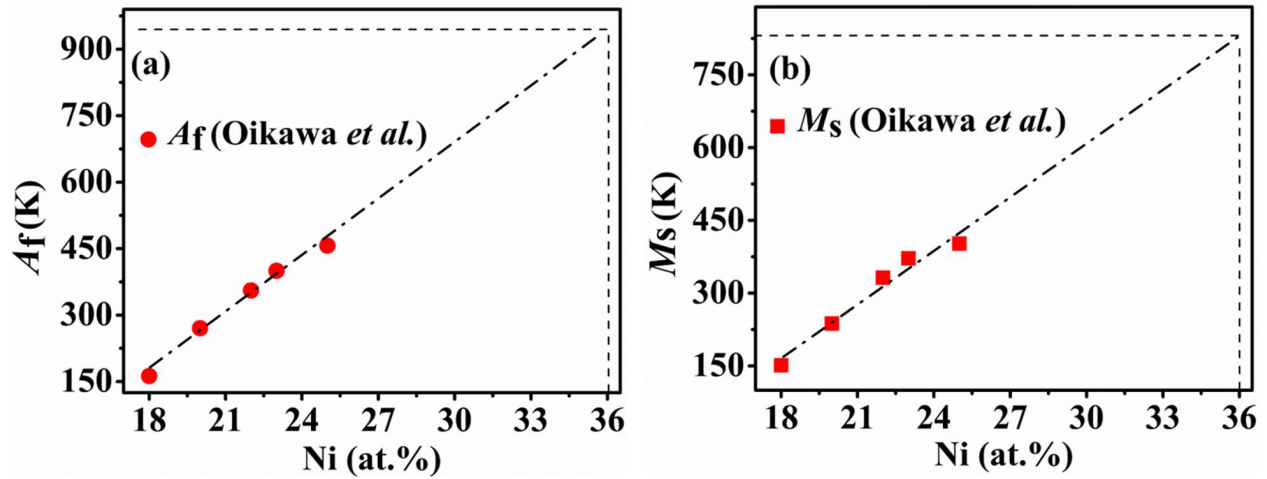

**Figure 3.** Variation of (a) austenite finish temperature ($A_f$) with Ni (at.%), and (b) martensite start temperature ($M_s$) with Ni (at.%) of Co-Ni-Ga alloy [after ref. 37]. Dotted lines represent extrapolated $A_f$, and $M_s$ values up to 36 at.% of Ni. corresponding its composition in CNG-1.

To assess the potential of the Co-Ni-Ga nanoparticles for shape memory applications, the martensitic transition has to be established. Thermocaloric technique such as differential scanning calorimetry (DSC) is widely employed for studying the first-order martensitic transition in shape memory alloys. However, the effectiveness of this technique diminishes when the bulk alloy is reduced to nanoparticles with a broad size distribution due to the reduced transformation enthalpy and wider transition temperature range [36]. It is difficult to observe such low enthalpy transitions through DSC curves as verified by us. Our search for an appropriate technique to observe the M ↔ A transition in these nanoparticles led us to temperature-dependent XRD technique as proposed by Fichtner *et al.* [28]. Temperature-dependent XRD, which is sensitive to changes in crystal structure of an SMA during the M ↔ A transition is better suited for this task. This technique also offers a distinct advantage by directly quantifying the evolution of A and M phases throughout the transition. Nevertheless, a significant limitation of temperature-dependent XRD is its inability to continuously monitor the structural changes occurring in the specimen with temperature, which makes precise determination of the transition temperatures difficult. Furthermore, the lack of recorded martensitic transition data of Co-Ni-Ga nanoparticles complicates the estimation of the transition temperature range in CNG-1 nanoparticles. In this regard, Oikawa *et al.*'s. [37] study on the effect of Ni at.% on austenite finish temperature ($A_f$) of bulk Co-Ni-Ga alloy has revealed that $A_f$ increases with increase in Ni at.%. though their studies were confined to alloys with up to 25 at.% Ni. As discussed in the previous section, in order to obtain single M phase in this CNG nanoparticles, the Ni content has to be 36 at.% or more. Since $A_f$ of Co-Ni-Ga alloy showed a nearly linear variation up to 25% Ni content, we extrapolated the data up to 36 at.% and found that $A_f$ increases to 946 K for this composition (c.f., figure 3 (a)). Armed with this estimate, we recorded XRD patterns of CNG-1 nanoparticles at 673 K and 1000 K on the heating cycle as shown in Figure 4 (a and b). At 673 K, a portion of the M phase (~59%) transformed into A phase (~ 41%). When the temperature was increased to 1000 K, the M phase completely transformed into A phase, as confirmed by the XRD pattern displayed in Figure 4(b). This clarifies that $A_f$ of CNG-1 sample is around 1000 K, which is close to the expected value for bulk Co-Ni-Ga alloy. Thus, the M→A transition in CNG-1 nanoparticles is established using structural data.

In a similar manner, martensitic start temperature ($M_s$) values of Oikawa *et. al* [37] were plotted and extrapolated up to 36 Ni at.% (c.f Figure 3(b)). This exercise showed that $M_s$ of CNG-1 is expected to be at ~ 830 K. To confirm this, CNG-1 was cooled down from 1000 K to 750 K and the XRD pattern was recorded (c.f., figure 4(c)). Figure 4(c) shows the appearance of M phase (6 %) along with A phase (94 %), confirming that $M_s$ > 750 K for CNG-1. Figure 4(d) shows that XRD pattern recorded when CNG-1 was cooled down to 300 K in which 100% M phase could be noticed. This establishes the A → M transition in CNG-1 nanoparticles. Thus, the two-way (A → M and M → A) shape memory effect is confirmed in CNG-1 nanoparticles. The percentages of A and M phases, along with their lattice parameters at various temperatures as determined through Rietveld refinement of the respective XRD pattern are listed in Table 3.

**Table 3.** Temperature-dependent structural properties of CNG-1 nanoparticles.

| Structural parameters | On heating cycle | | | On cooling cycle | |
|---|---|---|---|---|---|
| | **300 K** | **673 K** | **1000 K** | **750 K** | **300 K** |
| Percentage of phases | A = 0%<br>M = 100% | A = 41%<br>M = 59% | A = 100 %<br>M = 0% | A = 94%<br>M = 6% | A = 0 %<br>M = 100% |
| Lattice constants of M phase (Å) | $a$ = 2.709(3)<br>$c$ = 3.192(3) | $a$ = 2.714(5)<br>$c$ = 3.204(0) | --- | $a$ = 2.768(3)<br>$c$ = 3.136(6) | $a$ = 2.706(6)<br>$c$ = 3.199(8) |
| Lattice constant of A phase (Å) | --- | $a$ =5.751(8) | $a$ = 5.797(8) | $a$ =5.768(0) | --- |
| $\chi^2$ | 3.24 | 1.72 | 1.94 | 2.31 | 2.77 |

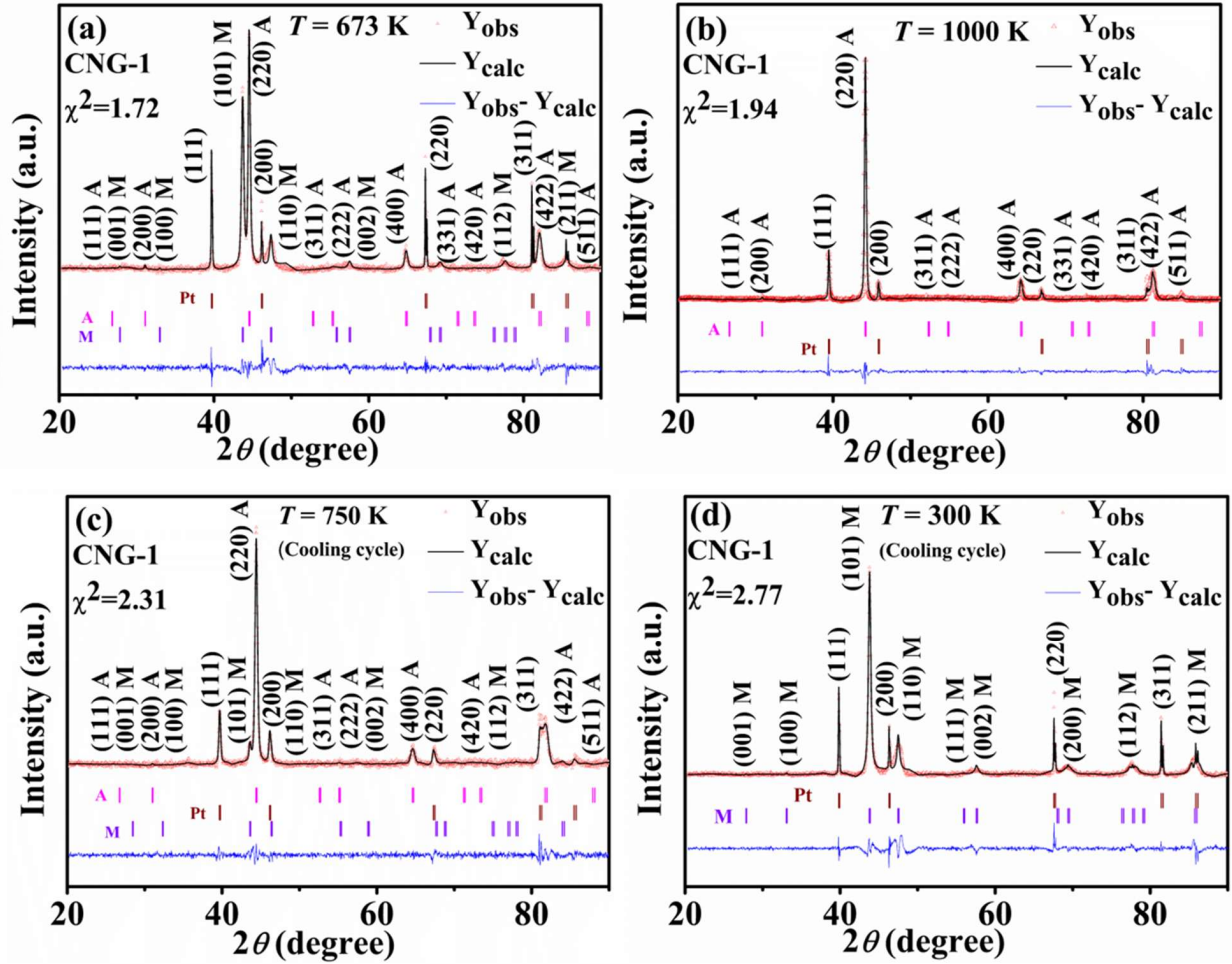

**Figure 4.** XRD patterns of CNG-1 nanoparticles recorded at (a) 673 K, (b) 1000 K during heating, and (c) 750 K and (d) 300 K during cooling.

To comprehend the influence of the γ-phase precipitate on the martensitic transition in Co-Ni-Ga nanoparticles, high-temperature XRD patterns were recorded for CNG-2 nanoparticles, which consist of approximately 81% M phase and 19% γ-phase at room temperature. When heated to 673 K, a portion of the M phase transforms into the A phase (29%) with a slight increase in γ phase (21%), as shown in Figure 5(a). As the temperature was further increased to 1000 K, the M phase completely transformed into A phase without any change in the percentage of γ phase (21%) (see figure 5(b)). The percentages of A, M, and γ phases, along with their lattice parameters at various high temperatures, determined through Rietveld refinement, are listed in Table 4. Upon cooling down to 300 K, the reappearance of 83% M and 17% γ phases confirms the two-way shape memory effect in CNG-2 nanoparticles (figure 5(d)). Thus, the γ phase does not contribute to the martensitic transformation and remains nearly constant across the temperature range.

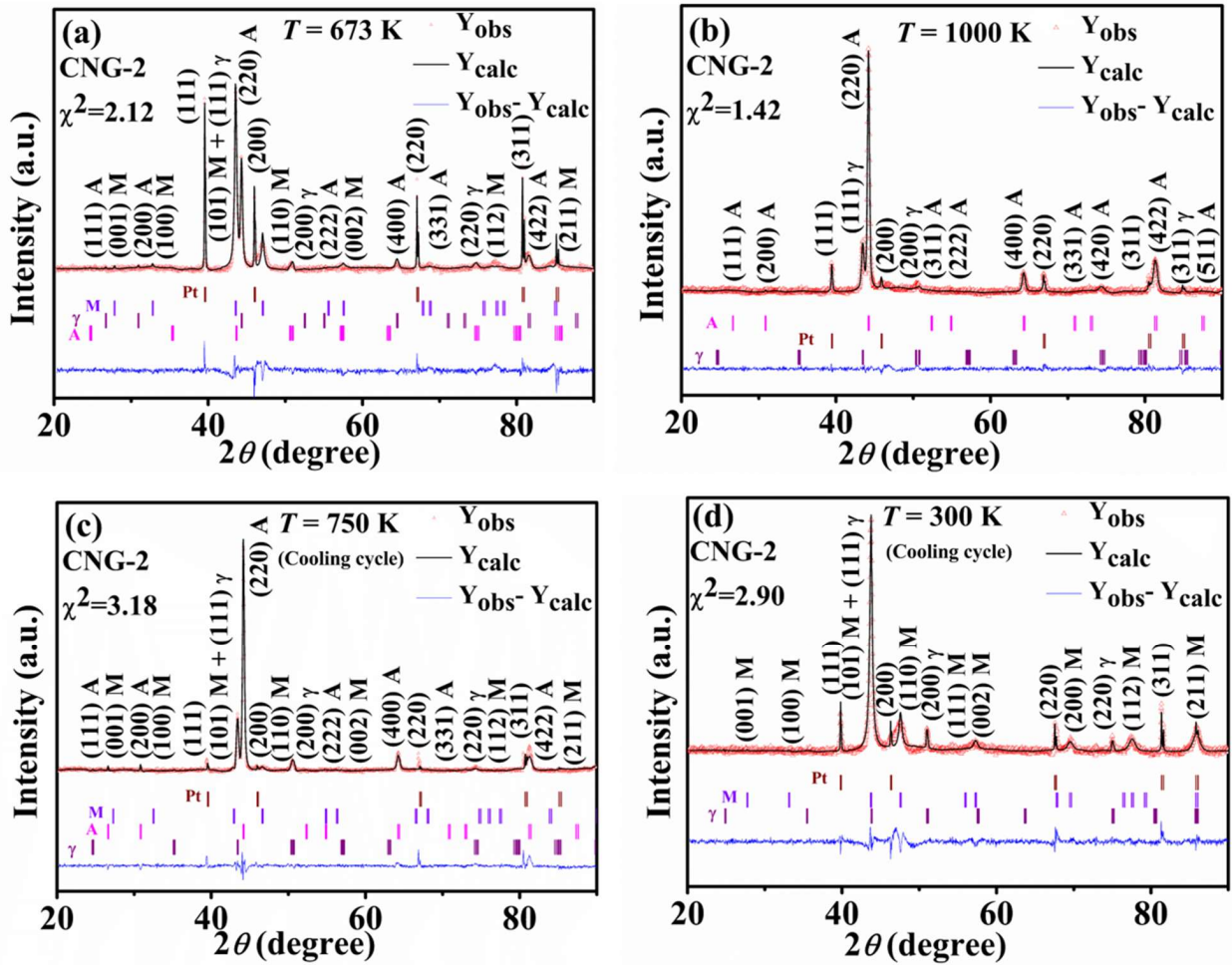

**Figure 5.** Rietveld refined XRD patterns of CNG-2 nanoparticles recorded at (a) 673 K, (b) 1000 K during heating, and (c) 750 K, (d) 300 K during cooling.

**Table 4.** Temperature-dependent structural properties of CNG-2 nanoparticles.

| Structural parameters | On heating cycle | | | On cooling cycle | |
|---|---|---|---|---|---|
| | 300 K | 673 K | 1000 K | 750 K | 300 K |
| Percentage of phases | M = 81%<br>γ = 19% | A = 28%<br>M = 51%<br>γ = 21% | A = 79 %<br>γ = 21% | A = 71%<br>M = 8%<br>γ = 21% | M = 83%<br>γ = 17% |
| Lattice constants of M phase (Å) | $a$ = 2.698(5)<br>$c$ = 3.216(8) | $a$ = 2.729(3)<br>$c$ = 3.203(3) | --- | $a$ = 2.754(6)<br>$c$ = 3.268(1) | $a$ = 2.703(0)<br>$c$ = 3.216(9) |
| Lattice constant of A phase (Å) | --- | $a$ = 5.779(8) | $a$ = 5.793(9) | $a$ = 5.797(6) | --- |
| Lattice constants of γ phase (Å) | $a$ = 3.574(9)<br>$c$ = 3.553(8) | $a$ = 3.586(4)<br>$c$ = 3.619(2) | $a$ = 3.596(9)<br>$c$ = 3.624(7) | $a$ = 3.603(5)<br>$c$ = 3.622(8) | $a$ = 3.580(6)<br>$c$ = 3.576(6) |
| $\chi^2$ | 2.06 | 2.12 | 1.42 | 3.18 | 2.90 |

To assess the potential of Co-Ni-Ga nanoparticles for nanoactuator applications at low temperatures, structural data were obtained on CNG-3 nanoparticles with A phase at room temperature. XRD patterns were recorded at 200 K, 100 K, and 50 K as shown in Figure 6. The XRD peaks around $2\theta$ of 43.5°, 51°, and 75° correspond to the FCC phase of the Cu sample holder. When CNG-3 was cooled to 200 K, some portion of the A phase (~37%) transformed into M phase. As the temperature is further lowered to 100 K, the M phase content increases to 67%. Further reducing the temperature to 50 K resulted in only a slight increase in the M phase content to ~69%. The percentages of the A and M phases, along with their lattice parameters at various low temperatures, were determined through Rietveld refinement and are listed in Table 5.

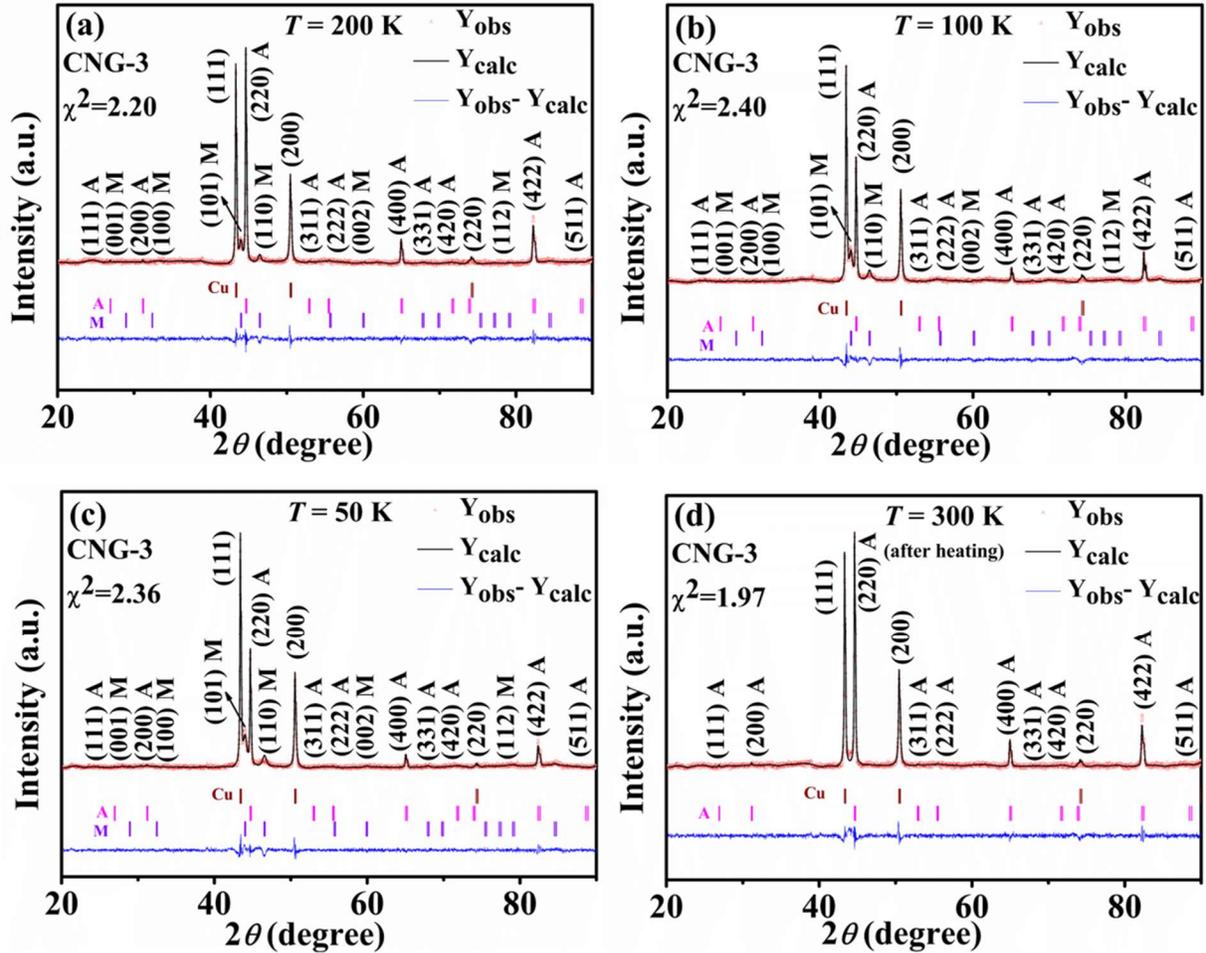

**Figure 6.** XRD patterns of CNG-3 nanoparticles recorded while cooling at (a) 200 K, (b) 100 K, and (c) to 50 K, and while heating back to (d) 300 K.

Interestingly, even at 50 K, only about 69% of the A phase transformed into M phase. From the previous discussion on CNG-1 and CNG-2 nanoparticles, it is evident that the martensitic transition occurs over a broad temperature range (approximately 500 K) in nanomaterials. This suggests that the remaining A phase in CNG-3 would continue to transform into A phase below 50 K. However, due to experimental constraints, our XRD measurements could be conducted at temperatures lower than 50 K. When CNG-3 was heated back to room temperature, the A phase was fully recovered in it as seen from figure 6(d). This not only established the two-way A ↔ M transition in CNG-3 nanoparticles but also demonstrated their utility in low-temperature SMA actuator applications.

**Table 5.** Temperature-dependent structural properties of chemically synthesized CNG-3 nanoparticles.

| Structural parameters | On cooling cycle | | | On heating cycle |
|---|---|---|---|---|
| | **200 K** | **100 K** | **50 K** | **300 K** |
| Percentage of phases | A = 63 %<br>M = 37% | A = 33 %<br>M = 67% | A = 31%<br>M = 69% | A = 100% |
| Lattice constants of M phase (Å) | $a$ = 2.764(3)<br>$c$ = 3.083(0) | $a$ = 2.763(8)<br>$c$ = 3.079(3) | $a$ = 2.758(7)<br>$c$ = 3.086(8) | --- |
| Lattice constant of A phase (Å) | $a$ = 5.737(5) | $a$ = 5.731(3) | $a$ = 5.730(2) | $a$ = 5.737(6) |
| $\chi^2$ | 2.20 | 2.40 | 2.36 | 1.97 |

3.4. Magnetic properties

Figure 7(a) illustrates the variation in magnetization with applied magnetic field ($M$-$H$ loops) of the Co-Ni-Ga nanoparticles recorded at 300 K. CNG-1 and CNG-2 nanoparticles exhibit soft ferromagnetic behavior, consistent with previous studies on Co-Ni-Ga nanoparticles [5, 12, 13, 28, 29]. The saturation magnetization ($M_{sat}$) at 300 K is 5.7 emu/g and 15.3 emu/g for CNG-1 and CNG-2, respectively. Notably, the CNG-2 nanoparticles dual (M + γ) phase displays the highest $M_{sat}$ value (15.3 emu/g) at 300 K among all three synthesized nanoparticles (see Table 6). The lower $M_{sat}$ of CNG-1 as compared to CNG-2 is attributed to a 5% reduction in Co content and the absence of the γ-phase. In contrast, CNG-3 nanoparticles exhibit very weak magnetic properties at room temperature with $M_{sat}$ of 2.9 emu/g at 4 Tesla. This substantial reduction in magnetic moment of CNG-3 is primarily due to its higher Ga atomic percentage and the presence of a single A phase. Thus, the concentration of the ferromagnetic elements and type(s) of crystalline phase(s) present in the alloy determine the magnetic moment of the alloys.

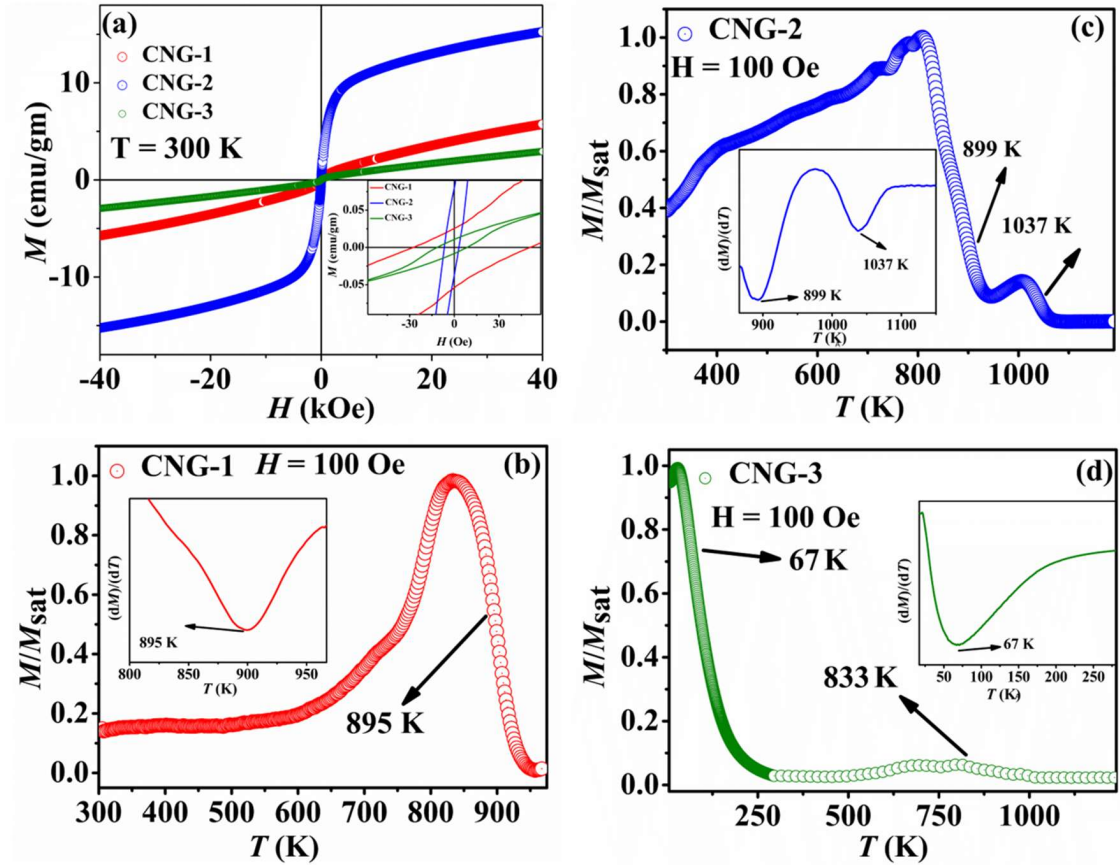

**Figure 7.** (a) Room-temperature *M-H* loops of CNG-1, CNG-2, and CNG-3 nanoparticles. *M-T* curves of (b) CNG-1, (c) CNG-2, and (d) CNG-3 nanoparticles.

Thermo-magnetization (*M-T*) curves of CNG-1, CNG-2, and CNG-3 nanoparticles were recorded under an applied field of 100 Oe (c.f., figure 7(b-d)). Figure 7(b) displays the *M-T* curve for CNG-1 nanoparticles. A closer examination of the *M-T* curve for CNG-1 reveals that the magnetic moment initially increases with temperature. This behavior occurs because CNG-1 nanoparticles exhibit martensitic phase with lower $M_{sat}$ at room temperature, which transforms into an A phase at higher temperatures. Therefore, the gradual increase in magnetic moment is attributed to the gradual formation of the A phase which has higher $M_{sat}$. A sudden drop in magnetization observed at $895 \pm 2$ K confirms the ferromagnetic-to-paramagnetic phase transition (Curie temperature ($T_C$)) of the A phase in CNG-1 nanoparticles. Similar behavior is seen in CNG-2 nanoparticles, with the $T_C$ of the A phase at $899 \pm 2$ K. Additionally, another ferromagnetic-to-paramagnetic phase transition occurs around $1037 \pm 2$ K, corresponding to the $T_C$ of the γ-phase present in the dual phase CNG-2 nanoparticles. Figure 7(d) shows the M-T curve of CNG-3 nanoparticles. These nanoparticles exhibit a ferromagnetic-to-paramagnetic phase transition at $67 \pm 2$ K, corresponding to the $T_C$ of the M phase, as a significant portion of the A phase converts to M at 50 K, as discussed earlier. As the temperature increases, another transition occurs around $833 \pm 2$ K. However, no such transition is observed in the cooling curve of CNG-3 (figure 8(c)) nanoparticles, indicating the irreversible structural transition around 833 K. When one compares the temperature dependent XRD patterns and the M-T curves of CNG-1, CNG-2 and CNG-3, it is evident that the $T_C$ of the M and A phases are higher than the martensite transition temperatures. This shows that these CNG nanoparticles are indeed FSMAs and can serve as magnetic field-induced actuators apart from being SMA based actuators.

**Table 6.** Room temperature magnetic properties of Co-Ni-Ga nanoparticles.

| Sample id | $M_{sat}$ (emu/g) | $H_c$ (Oe) | $M_r$ (emu/g) | $T_C$ (K) / phase |
|---|---|---|---|---|
| **CNG-1** | $5.7 \pm 0.1$ | 28 | 0.2 | $895 \pm 2$ / A |
| **CNG-2** | $15.3 \pm 0.3$ | 4 | 0.2 | $899 \pm 2$ / A |
|  |  |  |  | $1037 \pm 2$ / γ |
| **CNG-3** | $2.9 \pm 0.1$ | 10 | 0.1 | $67 \pm 2$ / M |

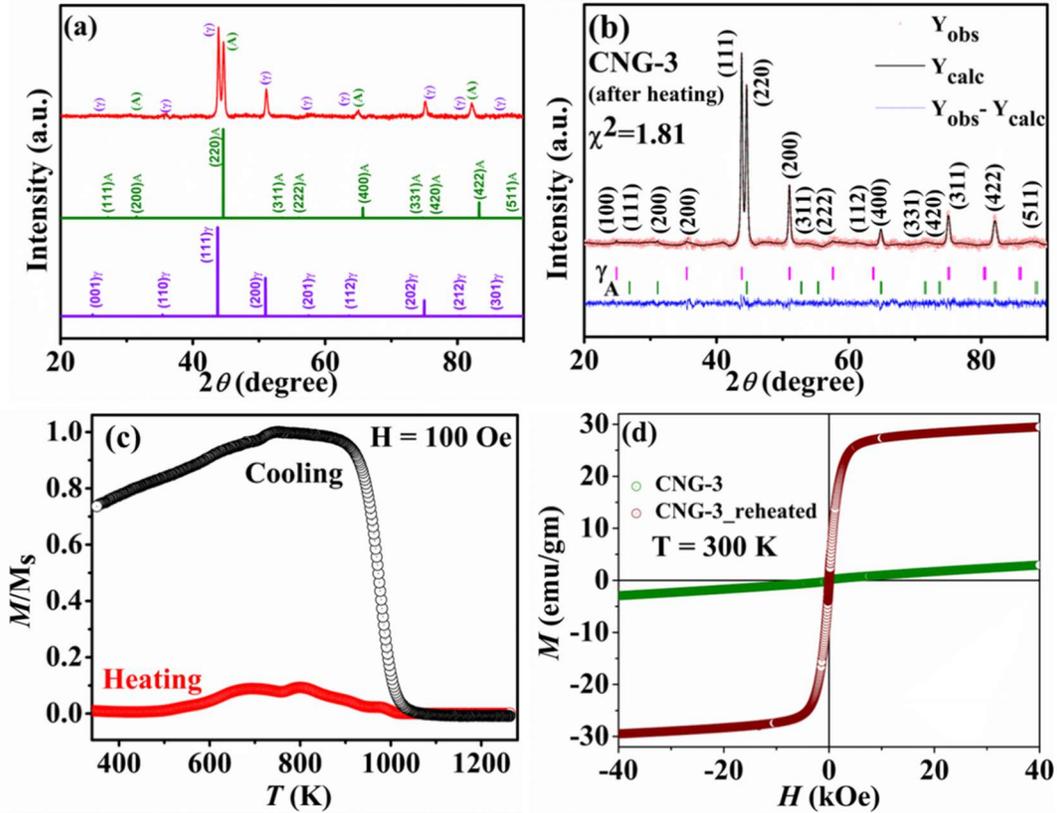

**Figure 8.** (a) Room-temperature XRD pattern of single A phase CNG-3 nanoparticle reheated to 1073 K along with (b) Rietveld refinement of the XRD pattern (c) temperature-dependent magnetic properties during heating and cooling (d) room temperature *M-H* loop of CNG-3 after cooling down from 1073 K.

To address the discrepancy in the *M-T* curve of CNG-3 nanoparticles, single A phase nanoparticles were heated to 1073 K. Figure 8(a, b) presents the room-temperature XRD pattern of CNG-3 nanoparticles after cooling from 1073 K. Significant amount of γ-phase (61%) is observed alongside the parent A phase. Therefore, the transition around 833 K is attributed to the irreversible transformation of the A phase into the γ-phase. Rietveld refinement was employed to confirm the dual A (39%) + γ (61%) phase and to determine the structural parameters. The lattice parameters of the A phase are $a = b = 5.751(2)$, whereas for the γ-phase, the values are $a = b =$ 3.578(4) and $c$ = 3.581(4). The emergence of the γ phase upon heating has also been observed in Heusler alloys [5,38].

To understand the substantial increase in the magnetic moment during the cooling curve compared to the heating curve (Figure 8(c)), a room-temperature *M-H* loop was recorded for the reheated CNG-3 nanoparticles. A significant increase in the magnetic moment is observed (Figure 8(d)), primarily due to the presence of a large amount of γ-phase. The γ-phase plays a key role in enhancing the magnetic properties, a phenomenon that has already been confirmed in the literature for Co-Ni-Ga nanoparticles [29]. Thus, the γ-phase can be introduced in controlled manner by adopting proper heat treatment conditions on the parent A phase of chemically synthesized CNG-3 nanoparticles.

## 4. CONCLUSION

Single M (designated as CNG-1), dual M + γ (designated as CNG-2), and single A phase (designated as CNG-3) Co-Ni-Ga nanoparticles have been synthesized using a template-free chemical route. Co-Ni-Ga nanoparticles with single M phase were obtained by increasing the Ni amount to 36 at.%, Co-Ni-Ga nanoparticles with dual (M + γ) phase could be achieved by increasing the Co amount to 41 at.%, and single A phase could be stabilized at room temperature in Ga rich (30 at.%) alloys. When heated to 1000 K, the M phase in CNG-1 nanoparticles transformed fully into A phase. When cooled to room temperature, single M phase was obtained, confirming the two-way martensitic transition in CNG-1 nanoparticles. It was found that the γ phase does not contribute/inhibit the martensitic transition in CNG-2. Cooling CNG-3 nanoparticles with room temperature A phase, resulted in only 69% conversion to M phase at 50 K. It was found that the γ phase could be controllably introduced in CNG-3 nanoparticles with single A phase at room temperature by heating it to 1073 K. The $T_C$ of the three Co-Ni-Ga nanoparticles is higher than their martensitic transition temperature, designating them as FSMAs. This study demonstrates that structural, magnetic, and ferromagnetic shape memory properties of Co-Ni-Ga nanoparticles can be tuned by choosing appropriate composition and heat treatment conditions depending upon the requirement of shape memory alloy application.